\documentstyle[12pt,epsf]{article}
\newlength{\mathspace}
\def\Dbar {\hbox{$D$\kern-0.52em\raise 0.2ex\hbox{/}\kern +0.1em}}
\def\dbar {\hbox{$\partial$\kern-0.52em\raise 0.2ex\hbox{/}\kern +0.1em}}

\begin{document}
\baselineskip=0.7cm
\setlength{\mathspace}{2.5mm}



\begin{titlepage}

    \begin{normalsize}
     \begin{flushright}
                
                 hep-th/9707058\\
     \end{flushright}
    \end{normalsize}
    \begin{LARGE}
       \vspace{1cm}
       \begin{center}
         {Supersymmetry and the Chiral Schwinger Model} 
       \end{center}
    \end{LARGE}

  \vspace{5mm}

\begin{center}
           
             \vspace{.5cm}

            Ricardo Amorim
           \footnote{E-mail address: amorim@if.ufrj.br}\\
Instituto de Fisica\\
Universidade Federal do Rio de Janeiro\\
RJ 21945-970, Caixa Postal 68528, Brazil\\
and \\
Ashok Das\\
Department of Physics and Astronomy\\
University of Rochester\\
Rochester, N.Y. 14627, USA\\

      \vspace{2.5cm}

    \begin{large} ABSTRACT \end{large}
        \par
\end{center}
 \begin{normalsize}
\ \ \ \
We have constructed the $N=\frac{1}{2}$ supersymmetric general Abelian model with asymmetric chiral couplings. This leads to a $N=\frac{1}{2}$ supersymmetrization of the Schwinger model. We show that the supersymmetric general model is plagued with problems of infrared divergence. Only the supersymmetric chiral Schwinger model is free from such problems and is dynamically equivalent to the chiral Schwinger model because of the peculiar structure of the $N=\frac{1}{2}$ multiplets.

\end{normalsize}

\end{titlepage}
\vfil\eject

\begin{large}
\noindent{\bf 1. Introduction:}
\end{large}

\vspace{.5cm}
Two dimensional soluble models such as the Schwinger model, the chiral Schwinger model, the massless Thirring  model, the gradient coupling model etc. have been studied extensively in the past [1-12]. All of these models can be described as interacting Abelian gauge theories. More recently, a general Abelian gauge theory with asymmetric vector and axial vector couplings was proposed [13], which is also soluble and reduces to various other models in different limits [14-15]. This general model can, therefore, be thought of as the parent theory for all the other soluble models.

Even though these models have been quite well studied, surprisingly, however, 
there does not exist a systematic study of the supersymmetric generalizations of
 these models. In fact, the only supersymmetric theory -- $N=1$ supersymmetric 
Schwinger model [16-17] -- that was constructed, already hinted at  problems in the supersymmetric  theory not present in the original theory [16]. Namely, although supersymmetry is known to lead to better ultraviolet behavior in theories, because the Schwinger model involves massless fermions and because supersymmetry introduces scalar partners to fermionic fields, the supersymmetric theory suffers from severe infrared divergence problems not present in the original theory. In this paper, we supersymmetrize the general model and study its properties systematically. In particular, we show that since the general model involves asymmetric chiral couplings, it does not allow for a $N=1$ supersymmetrization. However, the supersymmetric general model corresponds to a $N=\frac{1}{2}$ supersymmetrization which in a particular limit, leads  to a $N=\frac{1}{2}$ supersymmetric Schwinger model. The supersymmetric general model suffers from problems of infrared divergence much like the $N=1$ supersymmetric Schwinger mo
del. Surprisingly, however, the only model that is free from infrared divergence is the supersymmetric chiral Schwinger model which turns out to be dynamically equivalent to the chiral Schwinger model [7-9] itself because of the peculiar multiplet structures of $N=\frac{1}{2}$ supersymmetry. In section 2, we briefly review various ideas from $N=\frac{1}{2}$ supersymmetry including the structures of the multiplets as well as some simple theories. In section 3, we construct the $N=\frac{1}{2}$ supersymmetrization of the general model with asymmetric chiral couplings and derive from it the $N=\frac{1}{2}$ supersymmetric Schwinger model. We also indicate how the $N=1$ supersymmetric Schwinger model can be derived within this framework. In section 4, we discuss the problem of the infrared divergence and how the supersymmetric chiral Schwinger model is singled out to be free from such problems. We also point out how the supersymmetric chiral Schwinger model is dynamically equivalent to the chiral Schwinger model i
t

self. We present a brief conclusion in section 5.

\vspace{1cm}

\begin{large}
\noindent{\bf 2. $N=\frac{1}{2}$ Supersymmetry:}
\end{large}

\vspace{.5cm}
It is known that in $d=2$ (mod 8), we can define Majorana-Weyl spinors [18]. Consequently, the fundamental generator of supersymmetry in these dimensions can be a Majorana-Weyl spinor satisfying the algebra [19]

\begin{equation}
\left[Q_{\pm \alpha} , Q_{\pm \beta}\right]_{+} = \left(\frac{1\pm \gamma_{5}}{2} \gamma^{\mu} C\right)_{\alpha \beta} P_{\mu}
\end{equation}
where $Q_{\pm}$ represent Majorana-Weyl spinors ($\frac{1\pm \gamma_{5}}{2}$ define the right and the left handed projections) and $C$ is the matrix for charge conjugation. Theories providing representation of this symmetry algebra with only the right handed or the left handed charge are known as $N=\frac{1}{2}$ supersymmetric theories and are said to have $(\frac{1}{2}, 0)$ or $(0, \frac{1}{2})$ supersymmetry. By definition, such theories can involve chirally asymmetric interactions whereas the conventional $N=1$ supersymmetric theories are chirally symmetric and would correspond to the reducible $(\frac{1}{2},0) + (0,\frac{1}{2})$ supersymmetry.

The multiplet structure of $N=\frac{1}{2}$ supersymmetry is quite different from the conventional supersymmetry. Consequently, we describe, in this section, some simple $N=\frac{1}{2}$ supersymmetric theories before constructing the $N=\frac{1}{2}$ supersymmetrization of the general model. Although all of our discussion can be carried out in superspace, for simplicity and clarity, we would like to describe the theories in components. Furthermore, we would only restrict to theories with a supersymmetry generated by $Q_{-}$. Discussion of theories with supersymmetry generated by $Q_{+}$ is completely parallel.

Let us consider the free theory ($i=1,2,\cdots,n$)

\begin{equation}
{\cal L} = \frac{i}{2}\overline{\psi}_{+}^{i} \dbar \psi_{+}^{i} + \frac{1}{2}\partial_{\mu} A^{i} \partial^{\mu} A^{i}
\end{equation}
where $\psi_{+}^{i}$ is a right handed Majorana-Weyl spinor satisfying

\begin{equation}
\gamma_{5} \psi_{+}^{i} = \psi_{+}^{i}
\end{equation}
It is straightforward to show that this theory is invariant under the $N=\frac{1}{2}$ supersymmetry transformations

\begin{eqnarray}
\delta A^{i} & = & \overline{\epsilon}_{-} \psi_{+}^{i}\nonumber\\
\delta \psi_{+}^{i} & = & - i \dbar A^{i} \epsilon_{-}
\end{eqnarray}
where the parameter of transformation is a left handed Majorana-Weyl spinor satisfying

\begin{equation}
\gamma_{5} \epsilon_{-} = - \epsilon_{-}
\end{equation}

It is equally straightforward to check that the theory

\begin{equation}
{\cal L} = \frac{i}{2} \overline{\chi}_{-}^{i} \dbar \chi_{-}^{i} + \frac{1}{2} F^{i}F^{i}
\end{equation}
where $\chi_{-}^{i}$ represents a left handed Majorana-Weyl spinor is invariant under the supersymmetry transformations

\begin{eqnarray}
\delta \chi_{-}^{i} & = & F^{i} \epsilon_{-}\nonumber\\
\delta F^{i} & = & - i \overline{\epsilon}_{-} \dbar \chi_{-}^{i}
\end{eqnarray}
This brings out one of the differences from conventional supersymmetry in that $(\chi_{-}^{i}, F^{i})$ where $F^{i}$ represents an auxiliary field form a $N=\frac{1}{2}$ multiplet. Let us also note here that if we combine the two theories in eqs. (2) and (6), then,

\begin{equation}
{\cal L} = \frac{i}{2}\overline{\psi}_{+}^{i} \dbar \psi_{+}^{i} + \frac{i}{2} \overline{\chi}_{-}^{i} \dbar \chi_{-}^{i} + \frac{1}{2}\partial_{\mu} A^{i} \partial^{\mu} A^{i} + \frac{1}{2} F^{i} F^{i}
\end{equation}
is invariant under 

\begin{eqnarray}
\delta A^{i} & = & \overline{\epsilon}_{-} \psi_{+}^{i} + \overline{\epsilon}_{+} \chi_{-}^{i}\nonumber\\
\delta \psi_{+}^{i} & = & - i \dbar A^{i} \epsilon_{-} + F^{i} \epsilon_{+}\nonumber\\
\delta \chi_{-}^{i} & = & F^{i} \epsilon_{-} - i \dbar A^{i} \epsilon_{+}\nonumber\\
\delta F^{i} & = & - i \overline{\epsilon}_{-} \dbar \chi_{-}^{i} - i \overline{\epsilon}_{+} \dbar \psi_{+}^{i}
\end{eqnarray}
corresponding to the larger $N=1$ (or, equivalently,$(\frac{1}{2},0) + (0,\frac{1}{2})$) supersymmetry.

The structure of the gauge multiplet is also equally interesting. With the help of various two dimensional identities for the gamma matrices, it is easy to check that 

\begin{equation}
{\cal L} = - \frac{1}{4} F_{\mu \nu} F^{\mu \nu} + \frac{i}{2} \overline{\lambda}_{-} \dbar \lambda_{-}
\end{equation}
is invariant under the supersymmetry transformations

\begin{eqnarray}
\delta A_{\mu} & = & i \overline{\epsilon}_{-} \gamma_{\mu} \lambda_{-}\nonumber\\
\delta \lambda_{-} & = & -\frac{1}{2} \epsilon^{\mu \nu} F_{\mu \nu} \epsilon_{-}
\end{eqnarray}
The $N=\frac{1}{2}$ gauge multiplet is quite analogous to the scalar multiplet $(\chi_{-}^{i}, F^{i})$. In fact, comparing the structure of the theory as well as the transformations with those in eqs. (6) and (7), it is easy to identify $F \sim -\frac{1}{2} \epsilon^{\mu \nu} F_{\mu \nu}$. It follows now from our earlier discussion that the $N=1$ gauge theory would correspond to

\begin{equation}
{\cal L} = - \frac{1}{4} F^{\mu \nu} F_{\mu \nu} + \frac{i}{2} \overline{\lambda}_{-} \dbar \lambda_{-} + \frac{i}{2} \overline{\lambda}_{+} \dbar \lambda_{+} + \frac{1}{2} \partial_{\mu} M \partial^{\mu} M
\end{equation}
where $M$ represents a charge neutral scalar. In fact, it is easy to check that this theory is invariant under the larger supersymmetry transformations

\begin{eqnarray}
\delta A_{\mu} & = & i \overline{\epsilon}_{-} \gamma_{\mu} \lambda_{-} + i \overline{\epsilon}_{+} \gamma_{\mu} \lambda_{+}\nonumber\\
\delta \lambda_{-} & = & - \frac{1}{2} \epsilon^{\mu \nu} F_{\mu \nu} \epsilon_{-} - i \dbar M \epsilon_{+}\nonumber\\
\delta \lambda_{+} & = & - i \dbar M \epsilon_{-} - \frac{1}{2} \epsilon^{\mu \nu} F_{\mu \nu} \epsilon_{+}\nonumber\\
\delta M & = & \overline{\epsilon}_{-} \lambda_{+} + \overline{\epsilon}_{+} \lambda_{-}
\end{eqnarray}
It is worth noting here that the structure of the $N=1$ theory in two dimensions is different from that, say, in 4-dimensions in that it contains an additional scalar field $M$ and can, in fact, be thought of as a theory for a scalar multiplet. (Compare, for example, with eq. (8).)

\vspace{1cm}

\begin{large}
\noindent{\bf 3. $N=\frac{1}{2}$ Supersymmetric General Model:}
\end{large}

\vspace{.5cm}
In this section, we present the supersymmetrization of the general Abelian gauge theory with asymmetric chiral couplings [13-15]. We note that because of asymmetry in the chiral couplings, it is not possible to construct a $N=1$ supersymmetrization of this model. Consequently, we look for a $N=\frac{1}{2}$ supersymmetrization where the supersymmetry is generated by a left handed Majorana-Weyl spinor. The other case can also be constructed in a completely parallel manner.

Let us consider the theory described by the Lagrangian density

\begin{eqnarray}
{\cal L} & = & - \frac{1}{4} F^{\mu \nu} F_{\mu \nu} + \frac{i}{2} \overline{\lambda}_{-} \dbar \lambda_{-} + \frac{i}{2} \overline{\psi}_{+}^{i} (\Dbar^{(+)} \psi_{+})^{i} + \frac{1}{2} (D_{\mu}^{(+)} A)^{i} (D^{(+) \mu} A)^{i}\nonumber\\
 &  & + \frac{i}{2} \overline{\chi}_{-}^{i} (\Dbar^{(-)} \chi_{-})^{i} + \frac{1}{2} F^{i} F^{i} - e (1+r) \epsilon^{ij} A^{i} \overline{\lambda}_{-} \psi_{+}^{j}
\end{eqnarray}
where $(A^{i}, \psi_{+}^{i}, \chi_{-}^{i}, F^{i})$ with $i=1,2$ correspond to doublets under an internal $SO(2)$ symmetry. Furthermore, $(A^{i}, \psi_{+}^{i})$ carry a charge $e(1+r)$ while $(\chi_{-}^{i}, F^{i})$ carry a charge $e(1-r)$. Correspondingly, the covariant derivatives are defined to be

\begin{equation}
D_{\mu}^{(\pm) ij} =  \delta^{ij} \partial_{\mu} - e(1\pm r) \epsilon^{ij} A_{\mu}
\end{equation}
where $r$ measures the asymmetry in the couplings. With a little bit of algebra (including the use of Fierz identity in two dimensions), it is easy to check that the theory described by eq. (14) is invariant under the $N=\frac{1}{2}$ supersymmetry transformations

\begin{eqnarray}
\delta A^{i} & = & \overline{\epsilon}_{-} \psi_{+}^{i}\nonumber\\
\delta \psi_{+}^{i} & = & - i (\Dbar^{(+)} A)^{i} \epsilon_{-}\nonumber\\
\delta \chi_{-}^{i} & = & F^{i} \epsilon_{-}\nonumber\\
\delta F^{i} & = & - i \overline{\epsilon}_{-} (\Dbar^{(-)} \chi_{-})^{i}\nonumber\\
\delta A_{\mu} & = & i \overline{\epsilon}_{-} \gamma_{\mu} \lambda_{-}\nonumber\\
\delta \lambda_{-} & = & - \frac{1}{2} \epsilon^{\mu \nu} F_{\mu \nu} \epsilon_{-}
\end{eqnarray}
This is the most general supersymmetric Lagrangian that one can construct with the chirally asymmetric couplings. We could, of course, have included the multiplet $(\lambda_{+}, M)$ which corresponds to the right handed part of the gauge multiplet. However, they do not lead to any consistent interactions when the gauge couplings are asymmetric.

However, for  symmetric couplings, namely, when $r=0$,

\begin{equation}
D_{\mu}^{(+) ij} = D_{\mu}^{(-) ij} = D_{\mu}^{ij} = \delta^{ij} \partial_{\mu} - e \epsilon^{ij} A_{\mu}
\end{equation}
and eq. (14) provides the $N=\frac{1}{2}$ supersymmetrization of the Schwinger model. Furthermore, in this case, we can show that the Lagrangian density

\begin{equation}
{\cal L}' = \frac{i}{2} \overline{\lambda}_{+} \dbar \lambda_{+} + \frac{1}{2} \partial_{\mu} M \partial^{\mu} M + e \epsilon^{ij} (M \overline{\chi}_{-}^{i}\psi_{+}^{j} + M A^{i} F^{j} - A^{i} \overline{\lambda}_{+} \chi_{-}^{j})
\end{equation}
is independently invariant under the supersymmetry transformations of eq. (16) as well as

\begin{eqnarray}
\delta M & = & \overline{\epsilon}_{-} \lambda_{+}\nonumber\\
\delta \lambda_{+} & = & - i \dbar M \epsilon_{-}
\end{eqnarray}
The sum of the two Lagrangian densities, namely,

\begin{eqnarray}
{\cal L}_{TOT} & = & {\cal L} + {\cal L}'\nonumber\\
 & = & - \frac{1}{4} F^{\mu \nu} F_{\mu \nu} + \frac{i}{2} \overline{\lambda}_{-} \dbar \lambda_{-} + \frac{i}{2} \overline{\lambda}_{+} \dbar \lambda_{+} + \frac{1}{2} \partial_{\mu} M \partial^{\mu} M\nonumber\\
 &  & + \frac{i}{2} \overline{\psi}_{+}^{i} (\Dbar \psi_{+})^{i} + \frac{1}{2} (D_{\mu} A)^{i} (D^{\mu} A)^{i} + \frac{i}{2} \overline{\chi}_{-}^{i} (\Dbar \chi_{-})^{i} + \frac{1}{2} F^{i} F^{i}\nonumber\\
 &   & + e \epsilon^{ij} (M \overline{\chi}_{-}^{i} \psi_{+}^{j} + M A^{i} F^{j} - A^{i} \overline{\lambda}_{-} \psi_{+}^{j} - A^{i} \overline{\lambda}_{+} \chi_{-}^{j})
\end{eqnarray}
is, in fact, invariant under the larger supersymmetry

\begin{eqnarray}
\delta A^{i} & = & \overline{\epsilon}_{-} \psi_{+}^{i} + \overline{\epsilon}_{+} \chi_{-}^{i}\nonumber\\
\delta \psi_{+}^{i} & = & - i (\Dbar A)^{i} \epsilon_{-} + F^{i} \epsilon_{+}\nonumber\\
\delta \chi_{-}^{i} & = &  F^{i} \epsilon_{-} - i (\Dbar A)^{i} \epsilon_{+}\nonumber\\
\delta F^{i} & = & - i \overline{\epsilon}_{-} (\Dbar \chi_{-})^{i} - i \overline{\epsilon}_{+} (\Dbar \psi_{+})^{i}\nonumber\\
\delta A_{\mu} & = & i \overline{\epsilon}_{-} \gamma_{\mu} \lambda_{-} + i \overline{\epsilon}_{+} \gamma_{\mu} \lambda_{+}\nonumber\\
\delta \lambda_{-} & = & - \frac{1}{2} \epsilon^{\mu \nu} F_{\mu \nu} \epsilon_{-} - i \dbar M \epsilon_{+}\nonumber\\
\delta \lambda_{+} & = & - i \dbar M \epsilon_{-} - \frac{1}{2} \epsilon^{\mu \nu} F_{\mu \nu} \epsilon_{+}\nonumber\\
\delta M & = & \overline{\epsilon}_{-} \lambda_{+} + \overline{\epsilon}_{+} \lambda_{-}
\end{eqnarray}
This corresponds to the $N=1$ supersymmetric Schwinger model which was constructed earlier. Here, we have constructed it from the point of view of $N=\frac{1}{2}$ supersymmetry. 
\vspace{1cm}

\begin{large}
\noindent{\bf 4. Supersymmetric Chiral Schwinger Model:}
\end{large}

\vspace{.5cm}
From the structure of the $N=\frac{1}{2}$ supersymmetric general model, it is clear that there is a quartic coupling involving the scalar fields, $A^{i}$, and  the gauge field, $A_{\mu}$. This leads to self-energy graphs for the $A^{i}$ and the $A_{\mu}$ fields which are infrared divergent. In fact, it is easy to see that

{\bf figure caption is here}
\begin{equation}
\!\!\!\!\!\!
 e^{2}(1+r)^2\int\frac{d^2k}{k^2}
\end{equation}

Of course, we can add mass terms in a supersymmetric manner to cure the problem of infrared divergence. However, such a modification would correspond to starting theories which contain massive particles and are not soluble. (Incidentally, this is also the infrared divergence that was already noted [16] in the context of the supersymmetric Schwinger model corresponding to $r=0$.)

On the other hand, we see from eq. (22) that such infrared divergent terms would be absent for $r=-1$. In such a case, only the left handed fermions will have coupling to the gauge fields as is clear from eq. (14) and there would be no Yukawa coupling. In fact, the Lagrangian density (14), for $r=-1$ reduces to

\begin{eqnarray}
{\cal L} & = & - \frac{1}{4} F^{\mu \nu} F_{\mu \nu} + \frac{i}{2} \overline{\lambda}_{-} \dbar \lambda_{-} + \frac{i}{2} \overline{\psi}_{+}^{i} \dbar \psi_{+}^{i} + \frac{1}{2} \partial_{\mu} A^{i} \partial^{\mu} A^{i}\nonumber\\
 &  & + \frac{i}{2} \overline{\chi}_{-}^{i} \gamma^{\mu} (\delta^{ij} \partial_{\mu} - 2e \epsilon^{ij} A_{\mu}) \chi_{-}^{j} + \frac{1}{2} F^{i} F^{i}
\end{eqnarray}
Furthermore, since the multiplet $(A^{i}, \psi_{+}^{i})$ is noninteracting, we can take the dynamical theory to be

\begin{equation}
{\cal L}_{SCSM} = - \frac{1}{4} F^{\mu \nu} F_{\mu \nu} + \frac{i}{2} \overline{\lambda}_{-} \dbar \lambda_{-} + \frac{i}{2} \overline{\chi}_{-}^{i} \gamma^{\mu} (\delta^{ij} \partial_{\mu} - 2e \epsilon^{ij} A_{\mu}) \chi_{-}^{j} + \frac{1}{2} F^{i} F^{i}
\end{equation}
This would correspond to the supersymmetric chiral Schwinger model and surprisingly is free from the problem of infrared divergence. Furthermore, we note that this theory is really dynamically equivalent to the chiral Schwinger model since both $\lambda_{-}$ and $F^{i}$ are free fields. Consequently, all the results of the chiral Schwinger model would apply to its supersymmetric counterpart as well.
\vfil
\eject

\begin{large}
\noindent{\bf Conclusion:}
\end{large}

\vskip .5cm
We have constructed the $N=\frac{1}{2}$ supersymmetric general Abelian model with asymmetric chiral couplings in two dimensions. This leads to a $N=\frac{1}{2}$ supersymmetrization of the Schwinger model. We have shown that the supersymmetric general model is plagued with problems of infrared divergence. Only the supersymmetric chiral Schwinger model is free from infrared divergence. Dynamically, the supersymmetric chiral Schwinger model is equivalent to the chiral Schwinger model because of the peculiar $N=\frac{1}{2}$ multiplet structure.

A.D. would like to thank the members of the Instituto de Fisica at UFRJ and the theory group at Saha Institute for Nuclear Physics for hospitality where this work was carried out. He was supported in part by US DOE Grant No. DE-FG-02-91ER40685, NSF-INT-9602559 and by FAPESP. R.A. is supported by CNPq.
\vspace{1cm}

\begin{large}
\noindent{\bf References:}
\end{large}

\vskip .5cm
\begin{enumerate}
\item{ }J. Schwinger, Phys. Rev. 128 (1962)2425.
\item{ }W. Thirring, Ann. Phys. 3 (1958) 91.
\item{ } V. Glaser, Nuovo Cimento 9 (1958) 990; F. Scarf, Phys. Rev. 117 (1960) 868.
\item{ } K. Johnson, Nuovo Cimento 20 (1961) 773; C. Sommerfield, Ann. Phys. 26 (1964) 1.
\item{ }L. S. Brown, Nuovo Cimento 29 (1963) 617.
\item{ }C. R. Hagen, Nuovo Cimento 51B (1967) 169; {\it ibid} 51A (1967) 1033.
\item{ }C. R. Hagen, Ann. Phys. 81 (1973) 67; Phys. Rev. Lett. 55 (1985) 2223(C).
\item{ }R. Jackiw and R. Rajaraman, Phys. Rev. Lett. 54 (1985) 1219, 2060(E); {\it ibid} 55 (1985) 2224(C)2.
\item{ }A. Das, Phys. Rev. Lett. 55 (1985) 2126.
\item{ }R. Roskies and F. A. Schaposnik, Phys. Rev. D23 (1981) 558.
\item{ }A. Das and V. S. Mathur, Phys. Rev. D33 (1986) 489.
\item{ }A. Das and C. R. Hagen, Phys. Rev. D32 (1985) 2024.
\item{ }A. Bassetto, L. Griguolo and P. Zanca, Phys. Rev. D50 (1994) 1077.
\item{ }A. Das and M. Hott, Zeit. Physik C67 (1995) 707.
\item{ }C. R. Hagen, Univ. Rochester preprint, UR-1382.
\item{ }S. Ferrara, Lett. Nuovo Cimento 13 (1975) 629.
\item{ }A. Smailagic and J. A. Helayel-Neto, Mod. Phys. Lett. A2 (1987) 787.
\item{ }F. Gliozzi, J. Scherk and D. Olive, Nuc. Phys. B122 (1977) 253.
\item{ }M. Sakamoto, Phys. Lett. 151B (1985) 115.
\end{enumerate}

\vfil
\eject 

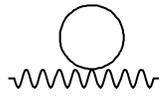
\end{document}